Comment on "Comparison of Holocene temperature reconstructions based on GISP2 multiple-gas-isotope measurements" by Döring and Leuenberger (2022).

Takuro Kobashi [a,*] and Tosiyuki Nakaegawa [b]

[a] Graduate School of Environmental Studies, Tohoku University, Sendai, Japan

[b] Meteorological Research Institute, Tsukuba, Japan

*Corresponding author. Email address: takuro.kobashi.e5@tohoku.ac.jp (T. Kobashi).

Döring and Lueneberger (Döring and Leuenberger, 2022) conducted an important work on Greenland temperature reconstruction from occluded air in an ice core for the Holocene by comparing various reconstructed temperatures. From their analyses, they conclude that their $\delta^{15}N$ temperature reconstruction ($\delta^{15}N$ method) is the best method among the compared methods, and Kobashi et al. temperature reconstruction with $\delta^{15}N_{excess}$ (Kobashi et al., 2017) should not be used for climate interpretation because $\delta^{40}Ar$ and $\delta^{15}N_{excess}$ data are not accurate enough. However, they failed to recognize important issues to reach their conclusions. Therefore, we comment on those points to make it clear that their conclusions are erroneous.

First, the firn densification models that Döring and Lueneberger used, namely Schwander et al. (Schwander et al., 1997) and Goujon et al. (Goujon et al., 2003), have never been validated for the process of firn densification in the time scale of decades owing to the lack of observational data. These models were constructed in the firn densification model proposed by Herron and Langway (Herron and Langway, 1980) with slight modification (Goujon et al., 2003; Schwander et al., 1997). The models were calibrated with present firn depth, temperatures, accumulation rates in the various fields in Greenland and Antarctica assuming an equilibrium state (Goujon et al., 2003; Schwander et al., 1997). These models produce slow changes of firn depth to multidecadal surface temperature changes (Kobashi et al., 2015), which have never been validated with actual observation. Because of this slow model response, these models were not able to reproduce a temperature history to satisfy both $\delta^{15}N$ and $\delta^{40}Ar$ and thus $\delta^{15}N_{excess}$ used for their analyses (Döring and Leuenberger, 2022). If the modelled firn depth can only change slowly on a multidecadal scale, most of observed multi-decadal $\delta^{15}N$ variability (such as Fig. 1) is explained by the changes in the temperature difference between the top and bottom of the firn. As the responses of $\delta^{40}Ar/4$ to temperature is smaller than those of $\delta^{15}N$ (Grachev and Severinghaus, 2003a, 2003b), expected $\delta^{40}Ar/4$ signals should be smaller than $\delta^{15}N$, which contradicts with the observed $\delta^{40}Ar$ signals and observed large variability of $\delta^{15}N_{excess}$ (=$\delta^{15}N$-$\delta^{40}Ar/4$). As they assumed that the model firn deification process is perfect, Döring and Lueneberger concluded that $\delta^{40}Ar/4$ data, calculated $\delta^{15}N_{excess}$, and the temperature reconstruction method using them are erroneous.

Kobashi et al. (Kobashi et al., 2010) have investigated other possibility that firn densification models cannot reproduce real "multi-decadal variation of firn depth" owing to the slow change of modelled firn densification in response to multi-decadal surface temperature change. Kobashi et al. developed a method to reconstruct surface temperature changes with minimum uses of the modelled firn densification process. The "integration" method calculates changes in temperature differences ($\Delta T$) between the top and bottom of firn (diffusive zone) using $\delta^{15}N$, $\delta^{40}Ar$, and $\delta^{15}N_{excess}$ (=$\delta^{15}N$-$\delta^{40}Ar/4$), and integrate the $\Delta T$ to calculate surface temperature coupling with a borehole temperature record (Kobashi et al., 2015, 2011, 2010). Holocene temperature was calculated by stabilizing the integration through fitting observed and modeled $\delta^{15}N$, coupled with borehole temperature (Kobashi et al., 2017). Although the calculation uses the same Goujon model (Goujon et al., 2003), it uses "temperature diffusion" process in the firn but not "firn densification" process (firn thickness variation has only second order influence on the temperature calculation in the Holocene). The temperature integration method using $\delta^{15}N_{excess}$ produces modeled $\delta^{15}N$ variation, which is larger than observed $\delta^{15}N$ as the modeled firn thickness can only vary slowly in a multidecadal to centennial scale. To be consistent with both observed $\delta^{15}N$ and $\delta^{15}N_{excess}$, the modelled firn thickness needs to respond faster. Then, a surface warming produces a temperature gradient in the firn (an increase in $\delta^{15}N$) and thinner firn thickness (a decrease in $\delta^{15}N$), which in total results in a smaller modeled $\delta^{15}N$ variability agreeing with the observation.

The model-data discrepancy cannot be solved by simply comparing firn models (Buizert et al., 2014; Goujon et al., 2003; Schwander et al., 1997) as the basic construct of these models are the same Herron and Langway model (Herron and Langway, 1980). We need to look outside of the box. For this purpose, the data for the past 1000 years including the longest observation from Ilulissat (which is highly correlated with the Summit temperature (Kobashi et al., 2017)) can provide useful tests (Fig. 1). For this purpose, dense ice core analyses (duplicate ice sample analyses in every 10 year) were made for GISP2 ice core (Kobashi et al., 2008) and similar analyses have been made from a neighboring core of NGRIP (Kobashi et al., 2015) (Fig. 1). The comparison of GISP2 and NGRIP reconstruction proves that the reconstructed Greenland temperature is real including the multi-decal variation. The trend is not a result of the post-coring fractionation on $\delta^{40}Ar$ as GISP2 and NGRIP bubble close-off depths are different between the two sites (artificial imprints should occur in different ages of $\delta^{40}Ar$ if any) (Kobashi et al., 2015). And the reconstructed temperatures are fully consistent with the borehole temperature records and other proxy temperature records (Kobashi et al., 2015), but not with Döring and Lueneberger's (Döring and Leuenberger, 2022).

In contrast, Döring and Leuenberger's reconstruction shows (Döring and Leuenberger, 2022) a little cooling for the past 1000 years "without" an observed warming in the early 20[th] century (the last data point of 1930). For the observed warming to be reconstructed by their method, observed $\delta^{15}N$ must have shown a rapid increase as firn thickness cannot respond to the rapid warming in the firn model. However, in reality, it was observed as only a small increase in $\delta^{15}N$ and a small decrease in $\delta^{40}Ar$, resulting in a rapid increase in $\delta^{15}N_{excess}$ (Fig. 1). This indicates that real firn thickness adjusted faster than the model calculates. As the integration method does not depend on the firn thickness changes, it correctly calculates the rapid warming in the early 20$^{th}$ century (Fig. 1). Possibly for the same reason, Döring and Leuenberger's reconstruction has smaller variability in a range of 1.9 °C over the past 1000 years in comparison to the observation of 2.5 °C and the reconstruction by the integration method of 3.1 °C as shown in Fig. 1.

In sum, Döring and Leuenberger failed to validate their reconstructed temperatures by comparing them with other independent temperature records, which is a mandatory process to be recognized as useful temperature proxies in the field of paleoclimatology. Also, they ignored the potential malfunctioning of firn deification models on a multi-decadal scale. In this comment, we provided evidence that $\delta^{15}N$ temperature reconstruction method underestimates temperature variability owing to the inability of the firn densification model to capture multidecadal variation. It is noted that most of the temperature reconstructions using inert gas in trapped air in ice cores uses only $\delta^{15}N$ coupled with firn densification models (e.g., (Huber et al., 2006; Kindler et al., 2014)) owing to the difficulty of measuring high precision $\delta^{40}Ar$. Thus, the issues discussed here are relevant to all of these results. Modeling of firn densification processes is critical not only for determining surface temperatures but also for calculating ages of trapped gasses, etc. This debate highlights the need to improving firn densification models for the better reconstruction of past climate.

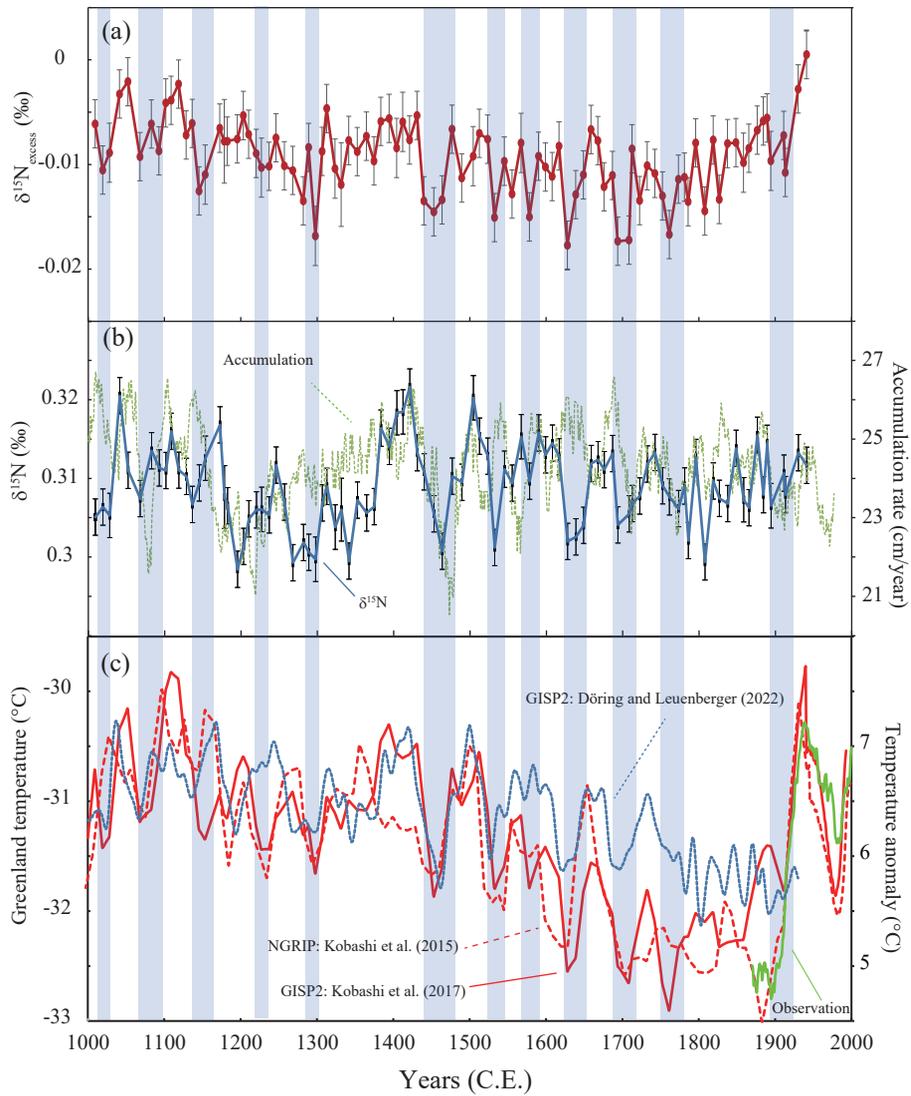

Figure 1. Greenland temperature reconstructions from trapped air in Greenland ice cores with input data. (a) GISP2 $\delta^{15}N_{excess}$, (b) GISP2 $\delta^{15}N$ and accumulation rate, (c) Greenland temperatures from GISP2, NGRIP, and observation at Ilulissat (31-year running means). Kobashi et al. reconstructions are on the scale of Greenland temperature (left) and Döring and Leuenberger is on the scale of Temerature anomaly (right). Observation has the same scale but different absolute temperature. The last data point for GISP2 isotopes is the year of 1941.